# Deep Learning-Based Detail Map Estimation for MultiSpectral Image Fusion in Remote Sensing

Arian Azarang, Nasser Kehtarnavaz

*Department of Electrical and Computer Engineering, University of Texas at Dallas, Richardson, USA*

Corresponding Author: Arian Azarang, E-mail: azarang@utdallas.edu

**Arian Azarang** received the BS degree and the first rank award from Shiraz University, Iran, in 2015, and the MS degree in electrical engineering from Tarbiat Modares University, Iran, in 2017. He is currently pursuing the PhD degree in electrical engineering at the University of Texas at Dallas. His research interests include signal and image processing, deep learning, remote sensing, and chaos theory. He has authored or co-authored 15 journal papers and conference papers in these areas. (**E-mail**: azarang@utdallas.edu)

**Nasser Kehtarnavaz** is an Erik Jonsson Distinguished Professor in the Department of Electrical and Computer Engineering and the Director of the Embedded Machine Learning Laboratory at the University of Texas at Dallas. His research interests include signal and image processing, machine learning, and real-time implementation on embedded processors. He has authored or co-authored ten books and more than 400 journal papers, conference papers, patents, manuals, and editorials in these areas. He is a Fellow of IEEE, a Fellow of SPIE, and a Licensed Professional Engineer. He is currently serving as Editor-in-Chief of Journal of Real-Time Image Processing. (**Email**: kehtar@utdallas.edu)

# Deep Learning-Based Detail Map Estimation for MultiSpectral Image Fusion in Remote Sensing

This paper presents a deep learning-based estimation of the intensity component of MultiSpectral bands by considering joint multiplication of the neighbouring spectral bands. This estimation is conducted as part of the component substitution approach for fusion of PANchromatic and MultiSpectral images in remote sensing. After computing the band dependent intensity components, a deep neural network is trained to learn the nonlinear relationship between a PAN image and its nonlinear intensity components. Low Resolution MultiSpectral bands are then fed into the trained network to obtain an estimate of High Resolution MultiSpectral bands. Experiments conducted on three datasets show that the developed deep learning-based estimation approach provides improved performance compared to the existing methods based on three objective metrics.

## 1. Introduction

In remote sensing, different types of images are captured via different sensors. Earth observation satellites normally collect spatial and spectral attributes of the earth surface in the form of so called PANchromatic (PAN) and MultiSpectral (MS) images. The fusion of or combining these two types of images, known as pansharpening, has been extensively studied in the literature (e.g., Lolli *et al.* 2017, Vivone *et al.* 2018, Vivone *et al.* 2014).

Recently, there has been an ever-growing use of deep learning in various image processing applications including in multispectral image fusion. Deep learning approaches of Convolutional AutoEncoder (CAE) and Denoising AutoEncoder (DAE) were utilized for pansharpening in (Azarang *et al.* 2019, Azarang and Ghassemian 2017). In (Huang *et al.* 2016), the DAE approach was considered by using a component substitution framework. In (Scarpa *et al.* 2018), a number of architectural and training variations to Convolutional Neural Network (CNN)-based image fusion was examined. In (Wei *et al.* 2017), a residual learning approach to form CNNs for pansharpening was discussed. A pansharpening method based on a deep neural network together with a new spectral loss were covered in (Eghbalian and Ghassemian, 2018). A good review of recent advances in remote sensing image fusion using deep learning architectures appears in (Liu *et al.*, 2018).

This paper presents an improvement of our previous work for estimation of detail maps in (Azarang, and Kehtarnavaz, 2020). The intensity component of the MS image is estimated by taking into consideration the joint multiplication of neighboring bands of a spectral band as part of the fusion approach known as component substitution. The nonlinear relationship between the estimated nonlinear intensity components and PAN image are modeled by using a deep neural network. After training the network, the LRMS image is fed into it to provide an estimation of the HRMS image. It is shown that this approach provides a more accurate estimation of the detail map for each spectral band, generating improved fusion outcomes.

The rest of the paper is organized as follows: Section 2 provides a description of the developed deep learning-based method. In Section 3, the datasets and the evaluation metrics used are stated followed by the experimental results in Section 4. Finally, the paper is concluded in Section 5.

## 2. Developed Deep Learning-Based Pansharpening Method

The general framework of the component substitution fusion can be expressed as follows:
$$\hat{\mathbf{M}}_k = \tilde{\mathbf{M}}_k + g_k(\mathbf{P} - \mathbf{I}) \qquad (1)$$

where $\hat{\mathbf{M}}_k$ and $\tilde{\mathbf{M}}_k$ denote High Resolution MS (HRMS) and Low Resolution MS (LRMS) images, respectively, $g_k$ the injection gain corresponding to the $k$-th spectral band, $\mathbf{P}$ the PAN image, and $\mathbf{I}$ the intensity image which is obtained via:

$$\mathbf{I}_k = \sum_{i=1}^{L} \omega_{i,k} \tilde{\mathbf{M}}_i \qquad (2)$$

where $L$ indicates the number of spectral bands covering the spectral signature of the PAN image, and $\omega_{i,k}$'s are the spectral weights. An illustration of the PAN and MS spectral spans for the three satellite sensors of GeoEye-1, QuickBird, and Pleiades-1A is depicted in Fig. 1 (https://www.satimagingcorp.com/).

Since, in practice, there is some overlap in the adjacent spectral bands, the detail map, $\mathbf{D}_k = \mathbf{P} - \mathbf{I}_k$, can be estimated more accurately for the four spectral bands shown in Fig. 1 as follows:

$$\mathbf{D}_1 = \mathbf{P} - \omega_{1,1}\tilde{\mathbf{M}}_1 - \omega_{1,2}\tilde{\mathbf{M}}_2 - \omega_{1,3}\tilde{\mathbf{M}}_3 - \omega_{1,4}\tilde{\mathbf{M}}_4 + b_{2,1}\tilde{\mathbf{M}}_1 \odot \tilde{\mathbf{M}}_2$$

$$\mathbf{D}_2 = \mathbf{P} - \omega_{2,1}\tilde{\mathbf{M}}_1 - \omega_{2,2}\tilde{\mathbf{M}}_2 - \omega_{2,3}\tilde{\mathbf{M}}_3 - \omega_{2,4}\tilde{\mathbf{M}}_4 + b_{1,2}\tilde{\mathbf{M}}_2 \odot \tilde{\mathbf{M}}_1$$
$$+ b_{3,2}\tilde{\mathbf{M}}_2 \odot \tilde{\mathbf{M}}_3$$

$$\mathbf{D}_3 = \mathbf{P} - \omega_{3,1}\tilde{\mathbf{M}}_1 - \omega_{3,2}\tilde{\mathbf{M}}_2 - \omega_{3,3}\tilde{\mathbf{M}}_3 - \omega_{3,4}\tilde{\mathbf{M}}_4 + b_{2,3}\tilde{\mathbf{M}}_3 \odot \tilde{\mathbf{M}}_2 \qquad (3)$$
$$+ b_{4,3}\tilde{\mathbf{M}}_3 \odot \tilde{\mathbf{M}}_4$$

$$\mathbf{D}_4 = \mathbf{P} - \omega_{4,1}\tilde{\mathbf{M}}_1 - \omega_{4,2}\tilde{\mathbf{M}}_2 - \omega_{4,3}\tilde{\mathbf{M}}_3 - \omega_{4,4}\tilde{\mathbf{M}}_4 + b_{3,4}\tilde{\mathbf{M}}_4 \odot \tilde{\mathbf{M}}_3$$

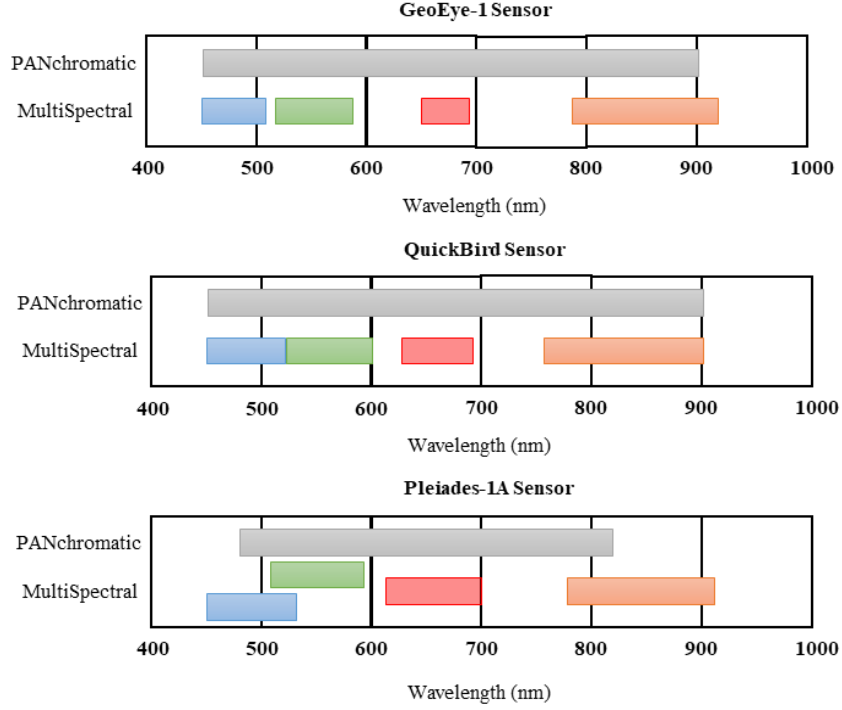

Fig. 1 Illustration of spectral characteristics for three satellite sensors (https://www.satimagingcorp.com/).

where $\odot$ denotes element-wise multiplication, and $b_{i,k}$'s are the weights for joint multiplication terms. In this equation, the joint multiplications of neighboring bands are considered in the computation of detail maps. The weights for the LRMS bands as well as the weights for the joint multiplication terms are obtained via the minimization of the Mean Squared Error (MSE) between the intensity component of a spectral band and its Modulation Transfer Function (MTF)-filtered PAN version. The detail map for each band is then obtained via Eq. (3). A flowchart of the fusion process is illustrated in Fig. 2.

The nonlinear intensity component for each spectral band and its corresponding histogram matched PAN image are partitioned into overlapping patches to serve as the input and the target of a DAE network as shown in Fig. 3. This network learns the nonlinear relationship between the intensity components of the spectral bands and its corresponding histogram matched PAN image. From a contextual point of view, the network attempts to inject the spatial information while keeping the spectral information. After the training phase or during the fusion operation, the Low Resolution MS (LRMS) bands are partitioned into overlapping patches and fed into the trained network to obtain an estimate of the HRMS image patches. Then, the patches are tiled back to obtain a fused image as follows:

$$\widehat{\mathbf{M}}_k = \overline{\mathbf{M}}_k + g_k(\mathbf{P}_k - \mathbf{I}_k) \tag{4}$$

where $\overline{\mathbf{M}}_k$ indicates the $k$-th tiled version of the estimated HRMS band, and the nonlinear intensity components $\mathbf{I}_k$'s are given by:

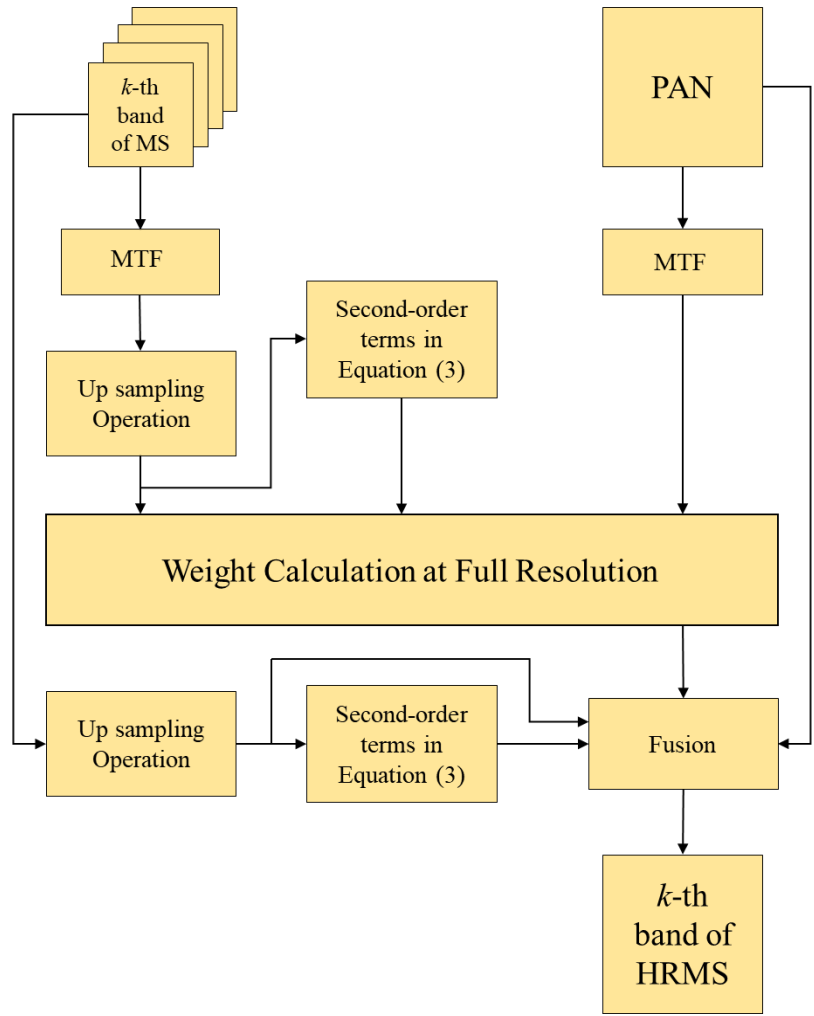

Fig. 2 Flowchart of the first part of the fusion process.

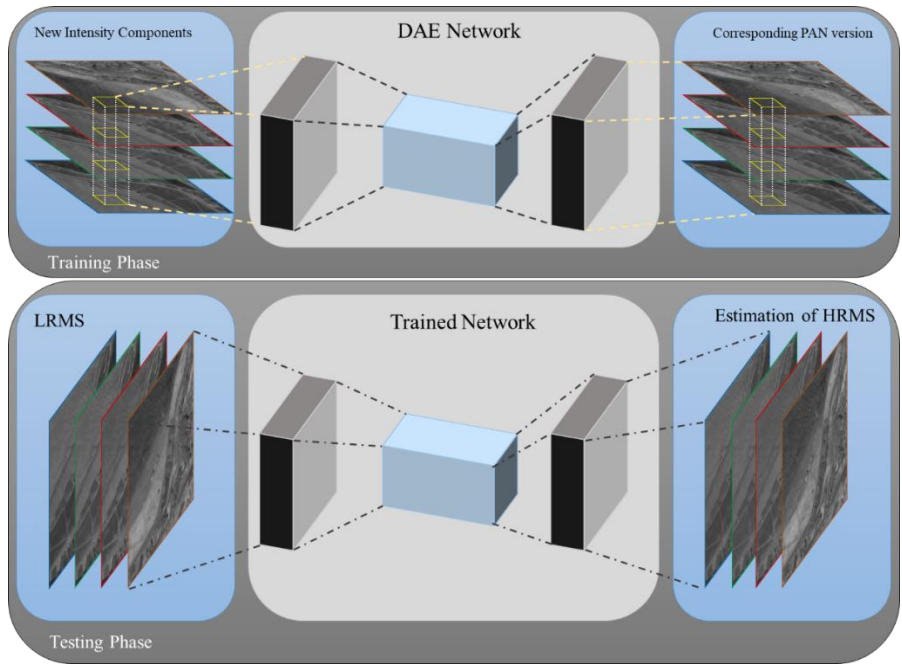

Fig. 3 Illustration of the Denoising AutoEncoder (DAE) network.

$$\mathbf{I}_1 = \omega_{1,1}\widetilde{\mathbf{M}}_1 + \omega_{1,2}\widetilde{\mathbf{M}}_2 + \omega_{1,3}\widetilde{\mathbf{M}}_3 + \omega_{1,4}\widetilde{\mathbf{M}}_4 + b_{2,1}\widetilde{\mathbf{M}}_1 \odot \widetilde{\mathbf{M}}_2$$

$$\mathbf{I}_2 = \omega_{2,1}\widetilde{\mathbf{M}}_1 + \omega_{2,2}\widetilde{\mathbf{M}}_2 + \omega_{2,3}\widetilde{\mathbf{M}}_3 + \omega_{2,4}\widetilde{\mathbf{M}}_4 + b_{1,2}\widetilde{\mathbf{M}}_2 \odot \widetilde{\mathbf{M}}_1 + b_{3,2}\widetilde{\mathbf{M}}_2 \odot \widetilde{\mathbf{M}}_3$$

$$\mathbf{I}_3 = \omega_{3,1}\widetilde{\mathbf{M}}_1 + \omega_{3,2}\widetilde{\mathbf{M}}_2 + \omega_{3,3}\widetilde{\mathbf{M}}_3 + \omega_{3,4}\widetilde{\mathbf{M}}_4 + b_{2,3}\widetilde{\mathbf{M}}_3 \odot \widetilde{\mathbf{M}}_2 + b_{4,3}\widetilde{\mathbf{M}}_3 \odot \widetilde{\mathbf{M}}_4 \quad (5)$$

$$\mathbf{I}_4 = \omega_{4,1}\widetilde{\mathbf{M}}_1 + \omega_{4,2}\widetilde{\mathbf{M}}_2 + \omega_{4,3}\widetilde{\mathbf{M}}_3 + \omega_{4,4}\widetilde{\mathbf{M}}_4 + b_{3,4}\widetilde{\mathbf{M}}_4 \odot \widetilde{\mathbf{M}}_3$$

Note that the detail injection gains of the intensity components are computed as described in (Vivone *et al.*, 2014) via the following equation

$$g_k = \frac{\text{cov}(\widetilde{\mathbf{M}}_k, \mathbf{I}_k)}{\text{var}(\mathbf{I}_k)} \quad (6)$$

where cov(**X**,**Y**) denotes the covariance between two images **X** and **Y**, and var(**X**) is the variance of **X**.

## 3. DATASETS AND EVALUATION METRICS

Three datasets associated with the GeoEye-1 (Washington, USA), QuickBird (Sundarbans, Bangladesh), and Pleiades-1A (Melbourne, Australia) sensors were examined to evaluate the developed deep learning-based method. The PAN and LRMS images for all the three datasets are of size 1024×1024 and 256×256 pixels, respectively. The developed method (labeled as IDM-DAE) is compared with some popular methods in the literature, namely AIHS (Rahmani *et al.*, 2010), GSA (Aiazzi *et al.*, 2007), GS2-GLP (Kallel, 2014), DNN (Huang *et al.*, 2015), MTF-GLP-HPM (Vivone *et al.*, 2013), BDSD (Garzelli *et al.*, 2007), FDIF (Azarang and Ghassemian, 2018), GLP-HRI (Vivone. 2019), and IDM – Base (Azarang and Kehtarnavaz, 2020).

To have an objective evaluation of the fusion outcomes, the full reference and no reference quality metrics are computed. In case of full reference metrics, Spectral Angle Mapper (SAM) (Vivone *et al.*, 2013) and Q4 (Vivone *et al.*, 2013) are computed. For no reference evaluations, $D_s$ (Palsson *et al.*, 2013) and $D_\lambda$ (Vitale, 2019) and the Quality of No Reference (QNR) (Vivone *et al.*, 2019) metrics are computed. The average values of these metrics are listed in Table I. $D_s$ and $D_\lambda$ reflect the spatial and spectral distortion, respectively, with their lower values indicating better performance. QNR combines $D_s$ and $D_\lambda$ in the form of an overall distortion. A description of the above objective metrics is provided next.

One way to measure spectral deviation of the fused product from the original MS image is to utilize the SAM metric, which is expressed as

$$SAM(\mathbf{A}_i, \mathbf{B}_i) = arccos\left(\frac{<\mathbf{A}_i, \mathbf{B}_i>}{\|\mathbf{A}_i\|\|\mathbf{B}_i\|}\right) \quad (7)$$

where **A** and **B** denote the fused and reference image, respectively, $<.,.>$ the inner product, and $\|.\|$ the vector $l_2$-norm. By averaging SAM values over all the pixels in an image, an overall SAM value can be acquired. The ideal value for the overall SAM is 0. In addition to SAM, the Q4 metric is also computed which is an extension of the Universal Image Quality Index (UIQI) (Wang and Bovik, 2002). This index is computed as follows:

$$\text{UIQI}(\mathbf{A}, \mathbf{B}) = \frac{\sigma_{AB}}{\sigma_A \sigma_B} \frac{2\mu_A \mu_B}{(\mu_A^2 + \mu_B^2)} \frac{2\sigma_A \sigma_B}{(\sigma_A^2 + \sigma_B^2)} \tag{8}$$

where $\sigma_{AB}$ is the covariance of **A** and **B**, and $\mu_A$ is the mean of **A**. Expressing Eq. (8) in vector format up to four bands is called Q4.

For the no reference quality case, the spectral and spatial distortions are quantified using $D_\lambda$ and $D_s$, respectively. $D_\lambda$ and $D_s$ are computed as follows:

$$D_\lambda = \sqrt[p]{\frac{1}{(N)(N-1)} \sum_{i=1}^{N} \sum_{j=1, i \neq j}^{N} \left|\text{UIQI}(\widehat{\mathbf{M}}_i, \widehat{\mathbf{M}}_j) - \text{UIQI}(\widetilde{\mathbf{M}}_i, \widetilde{\mathbf{M}}_j)\right|^P} \tag{9}$$

$$D_s = \sqrt[q]{\frac{1}{N} \sum_{i=1}^{N} \left|\text{UIQI}(\widehat{\mathbf{M}}_i, \mathbf{P}) - \text{UIQI}(\widetilde{\mathbf{M}}_i, \mathbf{P}^{\text{LP}})\right|^q} \tag{10}$$

where $\mathbf{P}^{\text{LP}}$ denotes the low resolution PAN image at the scale of MS image. It is worth mentioning that the optimal values for both $D_\lambda$ and $D_s$ is 0. The QNR metric combines these metrics together to provide a global average distortion as follows:

$$\text{QNR} = (1 - D_\lambda)(1 - D_s) \tag{11}$$

with its range being 0 to 1.

## 4. EXPERIMENTAL RESULTS

The fusion products at both reduced and full resolution are examined in this section. As seen from Table I, from an objective point of view, the improvement in the metrics are significant in comparison to the existing methods. This table consists of both the reduced and full resolution results. To see the difference in the reduced resolution, sample fusion products for each sensor are shown in Figs. 4 through 6. To see a comparison among the methods, the outcome corresponding to Pleiades-1A is shown in Fig. 6. The spatial superiority of the developed method is evident in Fig. 6(l). There are some regions in Fig. 6(d) and Fig. 6(h) that suffer from spectral distortion. The green area on the right side of the fused outcome is turned to light green for Fig. 6(c), Fig. 6(e), and Fig. 6(g). In Fig. (i) and Fig 6(j), the level detail injection is not sufficient. Our previous method generated similar results to those of the method developed in this paper except for some regions such as the overpass at the bottom of the images.

In terms of visual assessment at full resolution, sample fusion outcomes for each dataset are displayed in Figs. 7 through 9. In these figures, the arrows point to a visual difference. For instance, in Fig. 4, a part of the image is magnified for comparison purposes. It can be seen that the AIHS method suffers from color distortion especially in the green area. GS2-GLP has blurry outcome at the harbour edges. The green area in the FDIF image turned into the dark green in comparison to the green color of the LRMS image. The methods BDSD, GSA, and DNN oversharpened the outcome with a slight color distortion. The methods MTF-GLP-HPM, IDM – Base, and IDM-DAE produced similar outcomes in terms of visual perception but the developed method preserved the spectral information better than the MTF-GLP-HPM and its baseline methods.

Another example is shown in Fig. 9. In this figure, one can see that the methods AIHS and GSA suffered from the spectral distortion in few regions. The building edges when using the methods GS2-GLP, DNN and FDIF appeared blurred. The color information in the methods MTF-GLP-HPM and BDSD were lost in some regions. The method GLP-HRI suffered from oversharpening. The colors associated with the developed deep learning-based method (IDM-DAE) appeared better preserved across different datasets in comparison to its baseline version (IDM – Base).

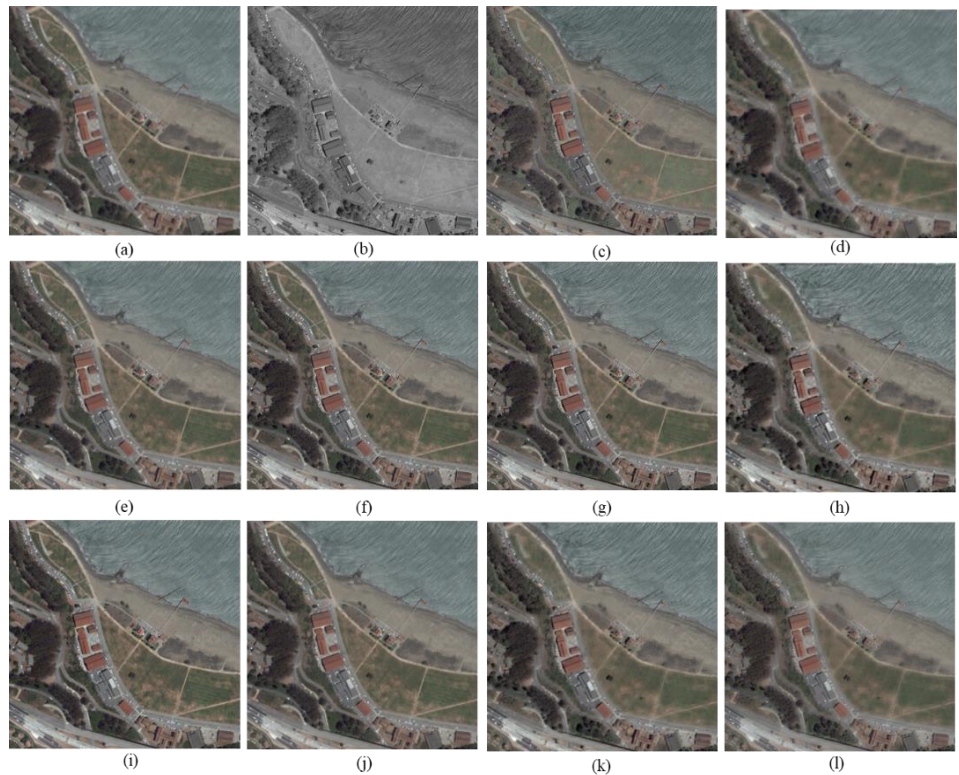

Fig. 4 Fusion outcomes of different methods for GeoEye-1 sensor at reduced scale: (a) LRMS, (b) PAN, (c) BDSD, (d) AIHS, (e) GSA, (f) GS2-GLP, (g) DNN, (h) MTF-GLP-HPM, (i) FDIF, (j) GLP-HRI, (k) IDM – Base, and (l) IDM – DAE.

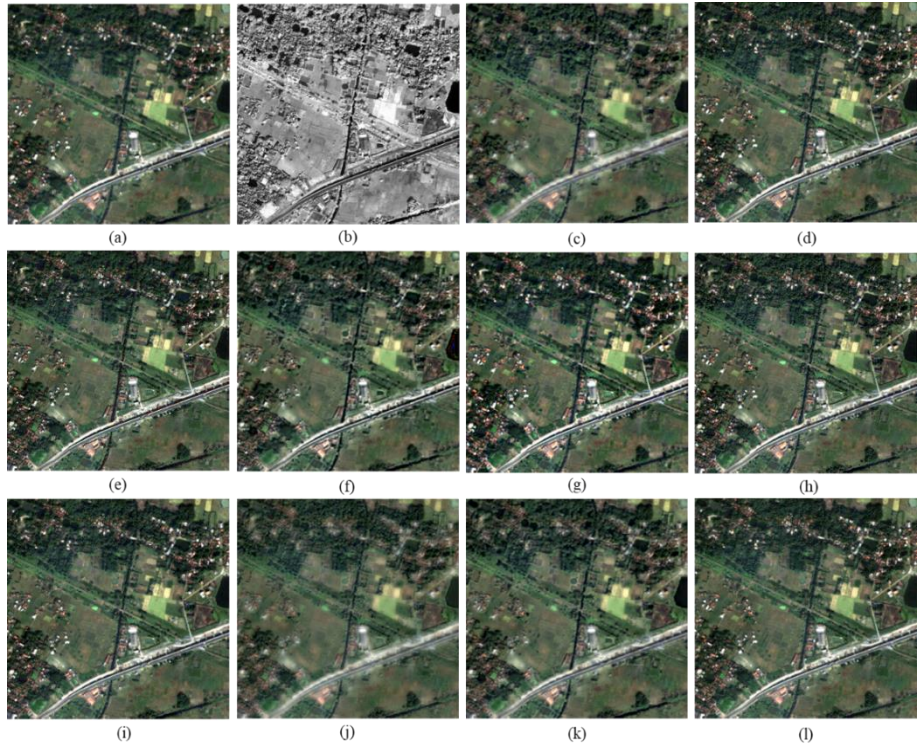

Fig. 5 Fusion outcomes of different methods for QuickBird sensor at reduced scale: (a) LRMS, (b) PAN, (c) BDSD, (d) AIHS, (e) GSA, (f) GS2-GLP, (g) DNN, (h) MTF-GLP-HPM, (i) FDIF, (j) GLP-HRI, (k) IDM – Base, and (l) IDM – DAE.

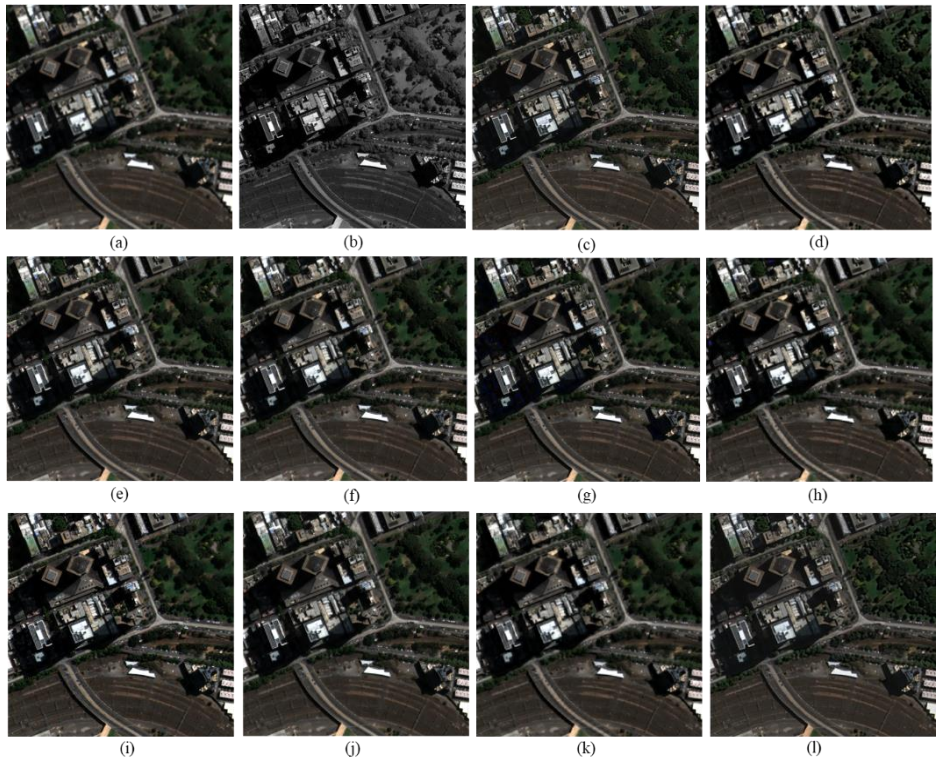

Fig. 6 Fusion outcomes of different methods for Pleiades-1A sensor at reduced scale: (a) LRMS, (b) PAN, (c) BDSD, (d) AIHS, (e) GSA, (f) GS2-GLP, (g) DNN, (h) MTF-GLP-HPM, (i) FDIF, (j) GLP-HRI, (k) IDM – Base, and (l) IDM – DAE.

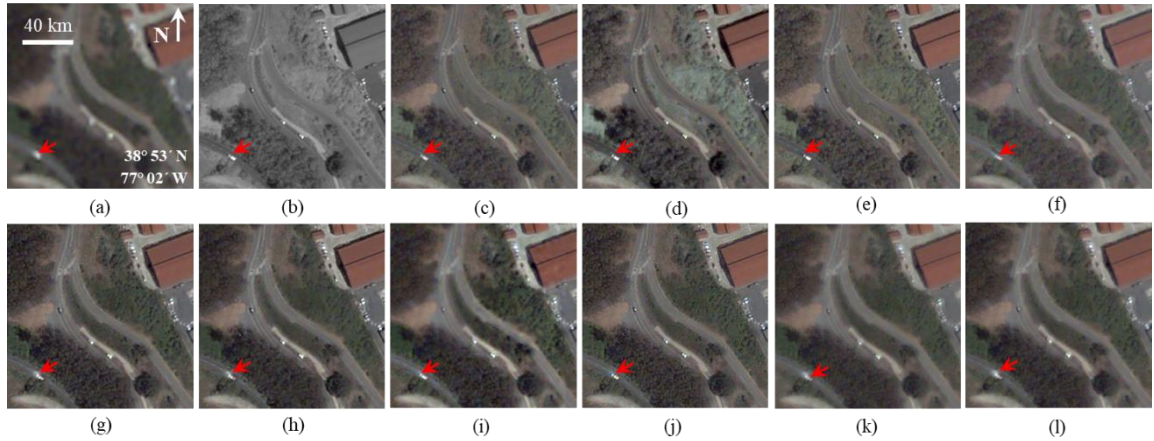

Fig. 7 Fusion outcomes of different methods for GeoEye-1 sensor at full scale: (a) LRMS, (b) PAN, (c) BDSD, (d) AIHS, (e) GSA, (f) GS2-GLP, (g) DNN, (h) MTF-GLP-HPM, (i) FDIF, (j) GLP-HRI, (k) IDM – Base, and (l) IDM – DAE.

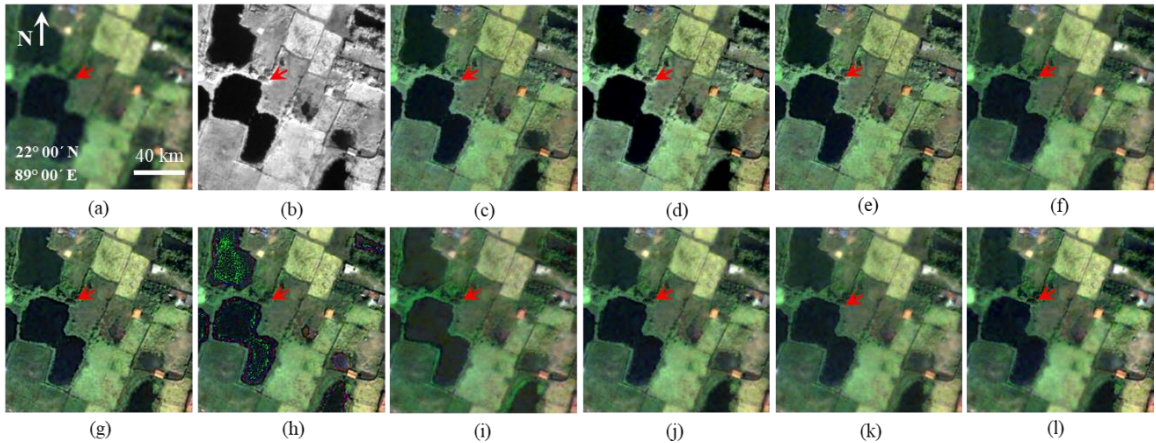

Fig. 8 Fusion outcomes of different methods for QuickBird sensor at full scale: (a) LRMS, (b) PAN, (c) BDSD, (d) AIHS, (e) GSA, (f) GS2-GLP, (g) DNN, (h) MTF-GLP-HPM, (i) FDIF, (j) GLP-HRI, (k) IDM – Base, and (l) IDM – DAE.

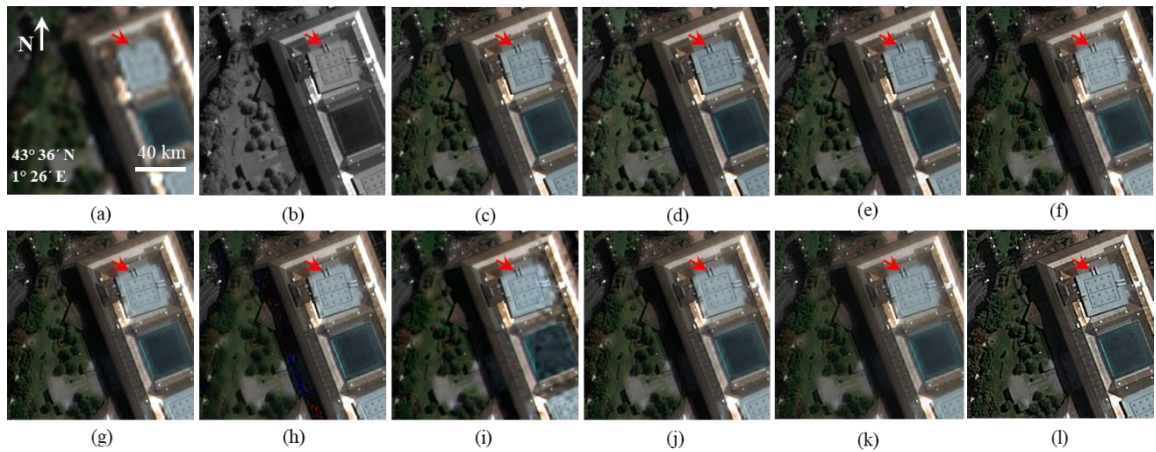

Fig. 9 Fusion outcomes of different methods for Pleiades-1A sensor at full scale: (a) LRMS, (b) PAN, (c) BDSD, (d) AIHS, (f) GSA, (e) GS2-GLP, (g) DNN, (h) MTF-GLP-HPM, (i) FDIF, (j) GLP-HRI, (k) IDM – Base, and (l) IDM – DAE.

TABLE I. Average Values of Objective Evaluation Metrics for Three Datasets

| | Reduced Resolution Analysis | | | | | | Full Resolution Analysis | | | | | | | | |
|---|---|---|---|---|---|---|---|---|---|---|---|---|---|---|---|
| | GeoEye-1 | | QuickBird | | Pleiades-1A | | GeoEye-1 | | | QuickBird | | | Pleiades-1A | | |
| Method | SAM | Q4 | SAM | Q4 | SAM | Q4 | $D_s$ | $D_\lambda$ | QNR | $D_s$ | $D_\lambda$ | QNR | $D_s$ | $D_\lambda$ | QNR |
| AIHS | 2.34 | 0.88 | 2.96 | 0.72 | 4.20 | 0.84 | 0.13 | 0.08 | 0.80 | 0.12 | 0.13 | 0.76 | 0.10 | 0.06 | 0.85 |
| GSA | 2.67 | 0.89 | 2.46 | 0.89 | 6.08 | 0.84 | 0.13 | 0.08 | 0.80 | 0.10 | 0.09 | 0.82 | 0.11 | 0.07 | 0.84 |
| GS2-GLP | 2.65 | 0.89 | 2.45 | 0.89 | 5.54 | 0.87 | 0.09 | 0.07 | 0.85 | 0.08 | 0.08 | 0.85 | 0.06 | 0.05 | 0.89 |
| DNN | 2.30 | 0.92 | 2.33 | 0.90 | 4.12 | 0.91 | 0.05 | 0.06 | 0.89 | 0.05 | 0.04 | 0.91 | 0.07 | 0.05 | 0.88 |
| MTF-GLP-HPM | 2.36 | 0.84 | **2.14** | 0.86 | **3.85** | 0.90 | 0.06 | 0.06 | 0.89 | 0.07 | 0.13 | 0.81 | **0.04** | 0.07 | 0.89 |
| BDSD | 2.76 | 0.88 | 2.53 | 0.87 | 5.23 | 0.87 | 0.07 | 0.09 | 0.84 | 0.08 | **0.03** | 0.89 | 0.10 | 0.06 | 0.84 |
| FDIF | 2.65 | 0.82 | 2.51 | 0.81 | 6.12 | 0.80 | 0.15 | 0.14 | 0.73 | 0.08 | 0.11 | 0.82 | 0.06 | 0.09 | 0.84 |
| GLP-HRI | 2.62 | 0.89 | 2.50 | 0.89 | 5.32 | 0.87 | 0.07 | 0.08 | 0.85 | 0.09 | 0.07 | 0.85 | 0.06 | 0.05 | 0.88 |
| IDM - Base | 2.31 | 0.91 | 2.35 | 0.90 | 4.54 | 0.91 | 0.06 | 0.06 | 0.89 | 0.06 | 0.05 | 0.90 | 0.09 | 0.07 | 0.85 |
| IDM - DAE | **2.12** | **0.95** | 2.25 | **0.91** | 4.07 | **0.93** | **0.04** | **0.04** | **0.92** | **0.03** | 0.04 | **0.93** | 0.07 | **0.05** | **0.90** |

## 5. CONCLUSION

A new band dependent detail injection pansharpening method has been developed in this paper by using a deep neural network. In this method, the computation of the intensity component of each MS band is carried out by considering the joint multiplications of neighboring bands in order to estimate the PAN image more accurately. The results obtained indicate the developed deep learning-based method outperforms the existing methods in terms of three objective metrics.